\documentclass{acm_sen_article}

\usepackage{lipsum}
\usepackage{flushend}
\usepackage{comment}
\usepackage{tcolorbox}
\usepackage{hyperref}
\usepackage{fontawesome5}

\newcommand{\resquest}[2]{
\begin{tcolorbox}[
    colback=blue!2,
    colframe=black!20,
    colbacktitle=blue!8,
    coltitle=black,
    title=\faQuestionCircle\ \textbf{#1},
    boxrule=0.3pt,
    arc=1mm
]
\textbf{#2}
\end{tcolorbox}
}

\begin{document}

\sloppy
\title{From Prompting to Engineering: A Research Agenda for Prompt Engineering in Software Engineering}

\numberofauthors{3} 

\author{
Vincenzo De Martino\textsuperscript{*}\\
    \affaddr{Universitat Politècnica de Catalunya}\\
    \affaddr{Barcelona, Spain}\\
    \email{vincenzo.de.martino@upc.edu}
\and
Giovanna Broccia\textsuperscript{*}\\
    \affaddr{CNR}\\
    \affaddr{Pisa, Italy}\\
    \email{giovanna.broccia@isti.cnr.it}
\and 
Fabiano Pecorelli\textsuperscript{*}\\
    \affaddr{Pegaso University}\\
    \affaddr{Naples, Italy}\\
    \email{fabiano.pecorelli@unipegaso.it}
\and 
Jennifer Horkoff\textsuperscript{*}\\
    \affaddr{University of Gothenburg and Chalmers University of Technology}\\
    \affaddr{Gothenburg, Sweden}\\
    \email{jennifer.horkoff@gu.se}
    \and 
Riccardo Coppola\\
\affaddr{Politecnico di Torino}\\
    \affaddr{Turin, Italy}\\
    \affaddr{riccardo.coppola@polito.it}
\and 
Antonino Ferraro\\
\affaddr{Pegaso University}\\
    \affaddr{Naples, Italy}\\
    \affaddr{antonino.ferraro@unipegaso.it}
\and 
Quim Motger\\
\affaddr{Universitat Politècnica de Catalunya}\\
    \affaddr{Barcelona, Spain}\\
    \email{joaquim.motger@upc.edu}
\and 
Emma McKenzie\\
\affaddr{University of Glasgow}\\
    \affaddr{Glasgow, Scotland}\\
    \affaddr{Emma.McKenzie@glasgow.ac.uk}
\and 
Shahbaz Siddeeq\\
\affaddr{Tampere University}\\
    \affaddr{Tampere, Finland}\\
    \affaddr{shahbaz.siddeeq@tuni.fi}
}

\maketitle
\begingroup
\renewcommand{\thefootnote}{*}
\footnotetext{Lead organizers.}
\endgroup
\begin{abstract}
Prompt engineering is increasingly used across Software Engineering (SE) activities, including requirements analysis, coding, testing, documentation, repository analysis, and planning. Yet prompts and related instruction artifacts are often created and evolved through task-specific and informal practices, with limited support for their systematic evaluation, management, traceability, and governance. To examine how SE can contribute to the maturation of these practices, we organized a structured community discussion at the First International Workshop on Empirical Prompt Engineering for Software Engineering (PROMPT-SE), co-located with EASE 2026. Participants discussed current prompting practices, challenges to their adoption and evaluation, and future directions for integrating prompt engineering into software development. We synthesized these discussions into five areas: prompt artifacts and standardization; evaluation and benchmarking; lifecycle integration; human–AI collaboration and skills; and governance, privacy, and technical debt. Based on these areas, we outline a research agenda to move prompt engineering from predominantly ad hoc interactions toward more systematic, maintainable, evaluable, traceable, and governable SE practices.
\end{abstract}


\section{Introduction}
Language Models (LMs), including both Large Language Models (LLMs) and Small Language Models (SLMs), are reshaping Software Engineering (SE), supporting activities such as requirements engineering, software development, testing, maintenance, software repository analysis, documentation, and empirical research \cite{hou2024large,10449667,11071447}. As these systems become increasingly integrated into the software development lifecycle, their effectiveness depends not only on model capabilities, but also on the artifacts used to specify, constrain, and guide their behavior.

Adapting LMs to specific SE tasks can involve different mechanisms, including fine-tuning, retrieval-augmented generation (RAG), and prompt engineering. While fine-tuning modifies model parameters and RAG augments generation with external knowledge, prompt engineering influences model behavior through instructions, context, examples, constraints, and interaction protocols. These prompt-related elements can affect the quality, reliability, reproducibility, cost, and maintainability of LM-assisted development activities.
However, despite its growing relevance, prompt engineering in SE remains largely informal and weakly integrated into established engineering processes. In many cases, prompts are created through individual experience, trial and error, or task-specific adaptation, with limited support for design, evaluation, reuse, traceability, and governance~\cite{chen2026promptware}.

This situation creates a critical need to shift from ad hoc prompting to prompt engineering as an explicit SE practice. Such a shift requires understanding how prompts are currently used across SE tasks, what challenges limit their reliable adoption, and how prompt-related artifacts can be managed as first-class engineering artifacts. It also raises broader questions about human-AI collaboration, specification-driven development, evaluation practices, privacy, compliance, technical debt, and the changing role of developers in LM-based workflows.

To contribute to this discussion, the \emph{First International Workshop on Empirical Prompt Engineering for Software Engineering (PROMPT-SE 2026)}~\footnote{https://conf.researchr.org/home/ease-2026/prompt-se-2026}, co-located with the International Conference on Evaluation and Assessment in Software Engineering (EASE 2026)~\footnote{https://conf.researchr.org/home/ease-2026}, provided a forum for researchers and practitioners interested in the empirical study of prompts and related instruction artifacts in SE. Alongside technical presentations, participants engaged in structured discussions about current practices, adoption and evaluation challenges, and possible future directions for prompt engineering in SE.

This article synthesizes the outcomes of these discussions and uses them to outline a community-informed research agenda for prompt engineering in SE. We organize the discussion into five focus areas: (1) prompt artifacts and standardization, (2) evaluation and benchmarking, (3) prompt engineering methods and workflow integration, (4) human–AI collaboration and skills, and (5) governance, privacy, and technical debt. Across these areas, we examine how established SE concepts and practices can contribute to moving prompt engineering from ad-hoc use toward a more systematic engineering discipline.
With this agenda, we aim to evolve prompt engineering from an informal interaction technique into a systematic, maintainable, evaluable, and governable SE practice.

The remainder of this article is organized as follows. Section \ref{sec:context} describes the workshop and discussion process. Sections \ref{sec:current_state} and \ref{sec:human} present the technical and human–organizational themes emerging from the discussion. Section \ref{sec:crosscutting} synthesizes their cross-cutting implications, and Section \ref{sec:agenda} presents the resulting research agenda.

\section{Workshop Context and Method}
\label{sec:context}
PROMPT-SE 2026 brought together researchers and practitioners interested in the empirical study of prompt engineering for SE. Beyond presenting ongoing research, the workshop aimed to identify recurring practices, challenges, and open questions that could inform future community efforts.

The workshop included technical presentations, a structured group activity, and a final plenary discussion. The accepted papers reflected the breadth of prompt engineering applications and concerns in SE. They investigated prompt-based approaches for conventional commit classification, goal extraction in Requirements Engineering, qualitative coding in empirical SE, and multi-agent code generation governed through TDD. Other contributions examined context-reset prompting as a strategy for reducing the energy cost of LLM-assisted SE and the limitations of generative AI for novice developers. Together, these studies covered not only different SE tasks, but also broader concerns around prompting strategies, evaluation, sustainability, governance, and human expertise. The technical program therefore provided concrete empirical and methodological perspectives that informed the subsequent discussion.

The structured activity involved six participants, who were divided into three groups of two. Each group examined a complementary perspective. The first focused on current practices in which prompting is used in SE, which practices participants considered effective, and what knowledge differentiates experienced from novice users. The second focused on challenges, including common failures, barriers to adoption, evaluation difficulties, and limitations to standardization. The third explored future directions, including possible changes to development practices, roles, tools, workflows, and human–AI collaboration. Given the small number of participants, the activity was intended to stimulate focused discussion rather than provide representative evidence of the broader SE community.

Each group was guided by a set of questions and asked to collect key observations. After the group discussions, representatives presented the main points in a plenary session. The organizers then consolidated the notes and synthesized the outcomes into thematic areas. While the discussion was exploratory, it provided a structured basis for identifying recurring concerns and research directions. The resulting agenda should therefore be interpreted as a community-informed synthesis rather than as an exhaustive empirical taxonomy.

Inspired by previous community-oriented research agendas in SE \cite{cruz2025greening,taibi2026research}, we synthesized the observations from the three discussion perspectives into five focus areas across two dimensions: technical---(i) prompt artifacts and standardization, (ii) evaluation and benchmarking, and (iii) lifecycle integration---and human and organizational---(iv) human--AI collaboration and skills and (v) governance, privacy, and technical debt.

\section{
Technical Dimensions}
\label{sec:current_state}
Prompt engineering research in SE spans heterogeneous activities. 
The PROMPT-SE technical program reflected this diversity, with studies addressing conventional commit classification, goal extraction in Requirements Engineering, qualitative coding of SE data, energy-aware prompting, novice developer support, and multi-agent code generation. These contributions illustrate that prompt 
engineering is being investigated well beyond code generation and across substantially different SE contexts. At the same time, the maturity and amount of available evidence vary across tasks and domains, making the coverage of prompt engineering research itself an important area for further investigation.

Across these applications, the workshop discussions converged on three recurring technical concerns: how prompt-related artifacts should be represented and managed, how prompt engineering should be evaluated, and how prompt artifacts can be operationalized within SE methods and workflows.

\subsection{Prompt Artifacts and Standardization}

Across the SE applications discussed at the workshop, participants consistently described the object being engineered as broader than the user prompt alone. Participants also identified contextual information, examples, constraints, templates, and interaction rules as relevant elements shaping model behavior. We use the term \emph{prompt artifact} to refer to the set of instructions and contextual elements deliberately assembled, stored, or reused to support an LM-based SE task.

\subsubsection*{From Prompts to Prompt Artifacts}

In SE, prompt artifacts can encode substantially different types of task-specific information. A code-generation artifact may include intended functionality, architectural constraints, coding conventions, testing expectations, and examples of desired outputs. Requirements analysis may additionally rely on stakeholder needs, domain assumptions, and previously elicited requirements, while repository analysis may incorporate project history, issue reports, or other organizational context. The engineering challenge therefore concerns not only how instructions are phrased, but also how the information required to support an LM-based task is selected, structured, and maintained.

Prompt artifacts can encode requirements, assumptions, domain terminology, organizational conventions, and quality expectations. They may also incorporate role- or domain-oriented guidance through mechanisms such as persona prompting, where the model is instructed to adopt the perspective of a specific stakeholder or expert role. Such mechanisms provide one way of making domain assumptions and task expectations more explicit within the interaction.

Workshop participants repeatedly associated effective prompting with the ability to provide relevant context and domain information. They also emphasized that prompting expertise alone may be insufficient when the user lacks knowledge of the application domain, project constraints, or organizational practices. These observations suggest that effective prompt engineering in SE may depend on the interaction among knowledge of LM behavior, domain expertise, and SE judgment. Understanding how these forms of knowledge contribute to prompt design and output quality remains an important empirical question.

\subsubsection*{The Need for Documentation and Standardization}

As prompt-based interactions become embedded in development workflows, their documentation becomes important for understanding and reproducing how generated artifacts were obtained. If prompt artifacts are not recorded, it becomes difficult to determine why a particular output was produced, how it relates to the original intent, or how the underlying instructions and context evolved over time. This limits reproducibility, traceability, and accountability.

Participants identified several potentially relevant elements to record, including the task goal, model and version, prompt text, system instructions, contextual information, examples, constraints, output format, evaluation criteria, and known limitations. For sensitive or high-impact tasks, additional metadata may include data sensitivity, privacy constraints, human-review requirements, and links to downstream artifacts.

Standardization, however, need not imply a single template for all prompt artifacts. SE tasks differ substantially, and requirements engineering, code generation, testing, documentation, and empirical research may require different structures and metadata. A more realistic objective is to establish shared documentation and reporting principles while allowing task-specific extensions. Such principles could make prompt artifacts easier to understand, reproduce, reuse, and compare.

\subsubsection*{Positioning Software Engineering}
SE can advance prompt engineering by treating prompts and related instruction artifacts as first-class engineering artifacts. Established SE practices such as requirements documentation, configuration management, design rationale, traceability, code review, and artifact versioning provide a foundation for systematic prompt management. From this perspective, prompt engineering should not be reduced to the search for better phrasings. Instead, it should involve the design, documentation, review, evolution, and reuse of artifacts that influence software development outcomes. By applying SE principles to prompt artifacts, the community can move from informal prompting towards practices that are more transparent, maintainable, and accountable.

\subsection{Evaluating Prompt Engineering in Software Engineering}

Evaluation emerged as one of the main challenges for the reliable adoption of prompt engineering in SE. Participants noted that teams often gain confidence in LM outputs through human expertise, manual review, testing, and informal evaluation. However, these practices are not always systematic, and they may vary significantly across tasks, teams, and organizations.

\subsubsection*{What Should Be Evaluated?}
A central difficulty is determining what should be evaluated. In prompt engineering, evaluation may target the prompt, the output, the interaction process, or the generated artifacts. A prompt may be well-structured but still produce poor output. Conversely, a prompt may produce a useful output once but fail to generalize across similar tasks, model versions, or contexts. This suggests distinguishing prompt characterization from its evaluation and considering four complementary levels. First, \emph{prompt characterization} concerns the systematic description of prompt artifacts and execution settings, such as prompt length, readability, number and type of few-shot examples, use of roles or personas, constraints, contextual information, and generation parameters such as token limits. Such characterization provides a basis for comparing prompting strategies and can support Automatic Prompt Engineering (APE) processes by making explicit the dimensions that can be varied or optimized. Second, \emph{artifact quality} concerns whether the prompt artifact adequately specifies the task, relevant context, constraints, and expected output. Third, \emph{outcome quality} concerns whether the resulting LM output or downstream software artifact satisfies task-specific quality criteria. Finally, \emph{process quality} concerns the effort, cost, number of iterations, reproducibility, and human intervention required to reach an acceptable outcome.

\subsubsection*{Multi-Dimensional Quality Criteria}
The discussions showed that prompt engineering cannot be evaluated using a single criterion. Correctness is important, especially for code generation, testing, and analysis tasks. However, participants also emphasized reliability, cost, latency, privacy, maintainability, reproducibility, sustainability, and trust. In many cases, the best prompt is not the one that produces the most fluent response, but the one that supports an acceptable trade-off among quality, effort, and risk. Non-determinism further complicates evaluation. The same prompt may produce different outputs across runs, models, or model versions. This instability makes it difficult to compare prompting strategies and to establish robust empirical evidence. Participants also mentioned overconfidence and hallucinations as particularly concerning failures, as LM outputs may appear plausible even when incorrect. In SE, such failures may propagate into code, tests, documentation, or design decisions if not properly reviewed. Another important aspect is cost. Prompt engineering often involves multiple iterations, and each iteration may consume time, computational resources, and financial resources. Cost is therefore not only an operational concern, but also an evaluation dimension. A prompt that produces slightly better outputs at a much higher cost may not be practical.

\subsubsection*{Benchmarking and Reproducibility}
Existing studies of prompt engineering in SE use heterogeneous tasks, models, datasets, prompts, metrics, and reporting practices, making cross-study comparison and cumulative knowledge building difficult. Participants emphasized the need for benchmarks and evaluation protocols that capture realistic SE scenarios. Reproducibility is particularly challenging because prompts depend on context. A reported prompt may not be sufficient to reproduce the result if the system prompt, conversation history, model version, temperature settings, retrieved documents, or hidden tool interactions are unavailable. This suggests that prompt engineering studies should report not only prompts but also relevant contextual and execution details. At the same time, benchmarks should avoid reducing prompt engineering to artificial tasks detached from practice. Real-world SE involves incomplete requirements, evolving constraints, organizational conventions, legacy code, and human judgment. Evaluation frameworks should combine controlled assessment with realistic scenarios and human-centered evaluation.

\subsubsection*{Positioning Software Engineering}

SE can contribute rigorous evaluation methodologies for prompt engineering by adapting practices from empirical SE, software testing, quality assurance, benchmarking, and software maintenance. In particular, SE can help define how prompt changes affect downstream artifacts, how prompt performance should be monitored across model versions, and how prompt-based workflows can be evaluated beyond one-shot accuracy. One direction discussed at the workshop was the systematic regression testing of prompt-based workflows. Such testing could assess whether changes to prompts, contextual information, model versions, or interaction protocols preserve previously established behavior. This raises a specifically SE-oriented challenge: defining what constitutes a regression in a non-deterministic LM-based workflow and which forms of behavioral variation should be considered acceptable. Developing suitable regression criteria, testing strategies, and monitoring mechanisms would be particularly relevant for organizations that rely on reusable prompt artifacts or prompt-based workflows in production settings.

\subsection{Prompt Engineering Methods and Workflows}
Workshop participants discussed how prompt artifacts can be used within different SE methods and activities, from requirements and planning to coding, testing, maintenance, and empirical research. This raises questions about where prompting can be applied, which prompting strategies are suitable for different types of tasks, and how to incorporate them into repeatable or partially automated workflows.

\subsubsection*{Prompting Across Software Engineering Activities}
Current uses of prompt engineering appear to vary by task. In coding, prompts are used to generate, explain, refactor, or debug code. In testing, prompts may support test generation, test case explanation, test repair, or fault localization. In documentation, prompts help summarize code, produce user-facing explanations, or improve technical writing. In requirements and planning, prompts support elicitation, prioritization, clarification, and task decomposition. In empirical research, prompts may support coding qualitative data, extracting information, or analyzing repositories. Despite this diversity, participants identified recurring practices across use cases. Effective prompts often include sufficient context, explicit constraints, examples, and clear expectations about the output. Negative constraints, such as instructions not to use certain solutions or not to make unsupported assumptions, were seen as useful mechanisms for controlling model behavior. However, participants also noted that such practices remain largely informal and are often learned through experimentation and trial and error. These differences raise the question of which prompting strategies are appropriate for different classes of SE tasks and how to incorporate them into repeatable, adaptable, or partially automated methods.

\subsubsection*{Toward Specification-Driven Development}

One of the strongest future-oriented themes concerned a possible shift toward more specification-driven development. Participants envisioned workflows in which developers increasingly express goals, constraints, requirements, and validation criteria that guide the generation of software artifacts. Under this view, LM-based development may shift part of the developer's effort from direct artifact production toward specification, validation, and decision-making, strengthening the connection between prompt engineering, requirements engineering, and software design.

This perspective also raises questions about stakeholder involvement and development practices. Faster generation of alternative implementations may support earlier validation and more interactive exploration of requirements and design choices. At the same time, lowering the barrier to artifact generation may increase the risk that software is produced without sufficient understanding of its quality, security, or maintainability implications. As one participant summarized this perspective, future development may involve ``less typing, more thinking.'' Whether such a shift actually occurs, for which tasks, and with what consequences for developers and stakeholders remain important empirical questions.

\subsubsection*{Traceability Across Prompts, Contexts, and Artifacts}
Lifecycle integration raises a traceability problem. When prompts and contextual information contribute to requirements, code, tests, or documentation, teams may need to recover how those artifacts were generated and which assumptions or constraints influenced them. Relevant trace links may therefore connect prompt artifacts with requirements, design decisions, generated outputs, tests, reviews, and subsequent revisions. Research is needed to determine which links provide practical value and how to maintain them without excessive overhead.

\subsubsection*{Positioning Software Engineering}
SE can advance prompt engineering by embedding prompting practices into lifecycle activities. Prompts should be connected to requirements, design decisions, tests, generated artifacts, and maintenance tasks. This would make prompt-based development more transparent and would support the long-term evolution of AI-assisted software systems. SE can also contribute process models for prompt-based development. Such models may define when prompts should be created, reviewed, evaluated, reused, or retired. They may also clarify how human validation should be integrated into prompt-based workflows and how prompt artifacts should be managed across teams.

\section{Human and Organizational Dimensions}
\label{sec:human}
Beyond technical considerations, participants emphasized that prompt engineering raises important human and organizational questions. As LM-based systems become increasingly embedded in development workflows, they influence how developers work, which skills are required, and how organizations govern AI-assisted practices. This section discusses the implications of prompt engineering for human-AI collaboration, education, governance, and long-term maintainability.

\subsection{Human-AI Collaboration, Roles, and Skills}

Participants discussed how LM-supported workflows may shift some development activities from direct artifact production toward specification, review, validation, and orchestration, raising questions about the skills and forms of judgment required from developers.

\subsubsection*{Changing Developer Roles}
In LM-supported workflows, some development activities may shift from direct artifact production toward specifying desired outcomes, providing context, constraining generation, and reviewing outputs. Participants therefore discussed a possible expansion of the developer role toward specification, validation, review, and orchestration. The extent of this shift is likely to depend on the task, organizational setting, and degree of automation, and remains an empirical question.
Such a shift may bring both benefits and risks. LM-based tools can reduce repetitive work, accelerate prototyping, and potentially allow developers to devote more attention to higher-level reasoning and validation. At the same time, increased reliance on generated artifacts may reduce developers' familiarity with the produced code, encourage over-reliance on model outputs, or lead to the acceptance of artifacts that are not fully understood. These risks may become particularly relevant during maintenance, debugging, and long-term system evolution.

\subsubsection*{Prompting Expertise and Domain Knowledge}
Participants emphasized that effective prompt engineering requires more than knowledge of prompting techniques. While AI-specific knowledge, such as zero-shot and few-shot prompting, may be useful, domain knowledge and organizational knowledge are often essential. A prompt can be syntactically well-structured but still ineffective if it lacks relevant domain assumptions, project constraints, or organizational conventions. These observations suggest that prompting expertise in SE may be inherently hybrid, combining knowledge of LM behavior with SE domain, and organizational knowledge. An important research question is therefore not simply which prompting techniques experts know, but which combinations of knowledge allow users to formulate effective instructions, recognize failure modes, and judge generated artifacts. 

\subsubsection*{Education and Prompting Literacy}
The changing role of developers raises important educational questions. If prompt-based development becomes a routine part of SE practice, education will need to address more than prompt formulation. Participants highlighted competencies such as task specification, context selection, constraint formulation, output evaluation, privacy awareness, and recognition of model limitations. Future work should investigate which of these competencies are genuinely new, which extend existing SE skills, and how they should be integrated into established topics such as requirements engineering, testing, quality assurance, security, and empirical evaluation.

\subsubsection*{Positioning Software Engineering}
SE can provide the educational and methodological foundation for responsible prompt engineering. SE education already emphasizes requirements, design, quality, testing, maintenance, and process. These areas can be extended to include prompt-based interaction with LM systems. Rather than treating prompt engineering as a separate skill, SE should teach how prompts participate in the production and evolution of software artifacts. This includes teaching students how to design prompts, evaluate outputs, manage risks, and understand the long-term consequences of AI-assisted development.

\subsection{Governance, Privacy, and Technical Debt}

At the organizational scale, prompt-based development raises questions about who may use particular models, what information can be shared with them, how interactions should be reviewed or recorded, and who remains accountable for the artifacts generated. Workshop discussions connected these questions with privacy, security, regulatory compliance, cost, and technical debt.

\subsubsection*{Privacy, Compliance, and Organizational Control}
Prompt-based workflows may expose source code, requirements, logs, or organizational knowledge to LM services, creating concerns around confidentiality, intellectual property, privacy, and regulatory compliance. Organizations may respond by restricting external services, adopting local models, or defining task-specific policies, each of which introduces different cost, capability, maintenance, and governance trade-offs. These concerns motivate governance mechanisms for prompt-based development. Such mechanisms may define which models can be used, what types of data can be included in prompts, which tasks require local execution, when human review is mandatory, and how prompt interactions should be logged or audited. Research is needed to understand which controls are effective, how they should vary across tasks and risk levels, and how they can be integrated into existing software governance processes.

\subsubsection*{Prompt-Induced Technical Debt}
Technical debt emerged as a particularly important concern. LM-based tools can accelerate development, but they may also introduce artifacts that developers do not fully understand, validate, or maintain. Participants noted that developers may become less familiar with generated code, rely on AI outputs during debugging, or accept solutions that work locally but create future maintenance costs. More broadly, participants linked prompt-based development to the risk of introducing maintenance costs when generated artifacts are insufficiently understood, validated, documented, or aligned with project constraints.

We use \emph{prompt-induced technical debt} as a provisional label for technical debt whose introduction or accumulation is influenced by the interaction among prompts, contextual information, generated artifacts, and human review. Such debt may arise from generated code, tests, documentation, requirements, configurations, or design decisions. It may result from underspecified prompts, missing context, weak evaluation, insufficient human review, or organizational pressure to prioritize speed over maintainability. This suggests a socio-technical research problem that requires both technical mechanisms and process-level governance.

\subsubsection*{Contextual Momentum and Overconfidence}
Participants also discussed situations in which an extended interaction continues to build on an earlier incorrect or incomplete assumption, producing subsequent outputs that remain coherent with the conversation but drift from the intended task. We provisionally refer to this observation as \emph{contextual momentum}. Rather than introducing it as an established phenomenon, we identify it as a candidate behavior for empirical investigation, including its relationship with conversational drift, error propagation, and user overreliance.

\subsubsection*{Positioning Software Engineering}
SE can contribute governance mechanisms for prompt-based development by adapting practices from configuration management, quality assurance, security engineering, compliance engineering, and technical debt management. Prompt engineering should be governed not only at the level of individual interactions, but also at the level of teams, organizations, and software lifecycles. In particular, SE can help define policies for prompt use, review processes for prompt artifacts, technical debt indicators for LM-generated artifacts, and audit mechanisms for high-impact prompt-based decisions. These contributions are essential to ensure that prompt-based workflows remain controllable, accountable, and maintainable over time.

\section{Cross-Cutting Reflections}
\label{sec:crosscutting}
Across the discussion, three cross-cutting themes emerged: standardization, evaluation, and lifecycle thinking. These themes connect the different focus areas and suggest broader directions for the evolution of prompt engineering in SE.

\subsection{Standardization}
Standardization was repeatedly identified as necessary for moving beyond ad-hoc prompting. This includes standardizing how prompts are documented, how prompt-based experiments are reported, how prompt templates are managed, and how organizations define safe prompting practices. However, standardization should not eliminate flexibility. Prompt engineering is highly context-dependent, and different SE tasks require different prompt structures and evaluation criteria. The challenge is  to define standards that provide enough structure to support reproducibility, traceability, and governance, while remaining adaptable to different tasks, domains, and organizational settings.

\subsection{Evaluation}

Evaluation is central to trustworthy prompt engineering. Participants emphasized that confidence in LM outputs often depends on human expertise, testing, and iterative validation. However, such practices must become more systematic. Future work should investigate quality models, benchmarks, human evaluation protocols, cost-aware metrics, and regression testing techniques for prompt-based workflows. Evaluation should consider the interaction among prompts, models, contexts, and outputs. A prompt that works well with one model may not work with another. A prompt that produces useful code may still introduce maintainability problems. A prompt that reduces development time may increase technical debt. Multidimensional evaluation is needed.

\subsection{Lifecycle Thinking}

Prompt engineering should not be evaluated only at the moment of interaction. Prompts can influence artifacts that persist, evolve, and affect future development activities. A generated code fragment may require maintenance. A generated test may shape future quality assurance. A generated requirement may influence design decisions. A generated explanation may become part of project documentation. This requires lifecycle thinking. Prompt engineering should be studied in relation to the artifacts it produces, the processes it affects, and the long-term consequences it introduces. Such thinking is central to SE and is one of the main ways it can contribute to the maturation of prompt engineering.

\section{Research Agenda}
\label{sec:agenda}
Based on the five themes synthesized from the workshop discussions, we formulate five research questions that capture open directions where SE research can contribute to the maturation of prompt engineering. These questions are intentionally broad and should be interpreted as a community-informed starting point rather than an exhaustive agenda.

\resquest{RQ1}{How should prompt artifacts be represented, managed, and evolved as first-class Software Engineering artifacts?}

Research should investigate which information constitutes a prompt artifact, which metadata are necessary across different SE tasks, and how prompt artifacts should be documented, versioned, reviewed, reused, and retired. A related challenge is establishing traceability among prompt artifacts, requirements, design decisions, code, tests, and other downstream artifacts.

\resquest{RQ2}{How should prompt engineering be evaluated in Software Engineering contexts?}

Research should investigate both the systematic characterization and evaluation of prompt engineering. Relevant directions include identifying meaningful prompt and execution features, evaluating artifact, outcome, and process quality, developing realistic benchmarks and prompt regression testing, and defining metrics for correctness, reliability, reproducibility, maintainability, cost, and human effort. Such characterization can also provide a basis for comparing prompting strategies and supporting APE.

\resquest{RQ3}{How can prompt artifacts be operationalized within Software Engineering methods and workflows?}

Research should investigate how prompt artifacts can be operationalized within SE methods and workflows. This includes understanding which prompting strategies are suitable for different types of SE activities, ranging from well-defined, repeatable tasks to exploratory, highly contextual ones. Particular attention should be given to methods that automate or partially automate the use and adaptation of prompt artifacts, including APE, prompt selection, and prompt refinement. Research should also examine how these methods interact with existing SE processes, how they should adapt as models and project contexts evolve, and under which conditions prompt engineering provides measurable benefits over alternative approaches.

\resquest{RQ4}{How does prompt engineering reshape human roles, skills, and human-AI collaboration?}

Research should examine how LM-supported workflows redistribute activities among specification, implementation, validation, and orchestration; which forms of domain, SE, and LM knowledge contribute to effective use; and whether prolonged reliance on generated artifacts affects developer understanding and judgment. These findings can inform the incorporation of prompt-related competencies into SE education and professional practice.

\resquest{RQ5}{How should prompt-based development be governed?}

Research should investigate how organizations govern model selection, sensitive information, human review, auditability, compliance, and accountability in prompt-based workflows. Another open direction concerns the relationship between prompting practices and technical debt, including whether prompt-induced technical debt can be identified, measured, and mitigated through existing or adapted SE mechanisms.

\section{Conclusion}
\label{sec:conclusion}
Prompt-based interaction with LMs is becoming increasingly relevant across SE activities, yet the engineering practices surrounding these interactions remain immature. Drawing on the discussions held at PROMPT-SE 2026, this article identified five areas where SE can contribute: (1) prompt artifacts and standardization, (2) evaluation and benchmarking, (3) prompt engineering methods and workflow integration, (4) human–AI collaboration and skills, and (5) governance and technical debt.

The agenda is not intended as an exhaustive taxonomy, but as a community-informed starting point. Its central argument is that progress in prompt engineering requires moving beyond optimizing individual prompt formulations. Prompts and their surrounding context increasingly participate in the specification, production, and evolution of software artifacts and should therefore be studied with the same concerns for evaluation, traceability, maintainability, and governance that SE applies to other development artifacts. Moving from prompting to engineering means understanding not only how to obtain better outputs, but how to build prompt-based practices that can be evaluated, evolved, and trusted over time.


\section*{Acknowledgements}

The authors thank all the authors who submitted their work to PROMPT-SE 2026 and all participants for contributing to the discussions throughout the workshop. This work has been partially supported by Grant PID2024-156019OB-I00 funded by MICIU/AEI/10.13039/501100011033 and the European Regional Development Fund (ERDF), European Union. This work was also supported by the project \emph{``Integrazione di Modelli Linguistici di Grandi Dimensioni (LLM) Open Source e Recupero di Informazioni (RAG) per la Formazione Avanzata (MIRIAM)''} (CUP/ID: FRC2024012), funded under the \emph{FRC 2024 -- Progetti di Ricerca di Ateneo} of Università Pegaso (Decree No. 1072, November 18, 2024). Generative AI (ChatGPT) was used exclusively to improve the language and readability of this manuscript. All authors reviewed, edited, and validated the final content and take full responsibility for the manuscript.

\bibliographystyle{plain}
\bibliography{biblio}

\end{document}